\documentclass{rspublic}

\input epsf
\let\sec=\section
\let\ssec=\subsection

\def\japitem#1{\smallskip\noindent\rlap{#1}\hglue1.8em\hangindent1.8em}

\def\japref{\parskip=0pt\par\noindent\hangindent\parindent
    \parskip =2ex plus .5ex minus .1ex}

\def\gs{\mathrel{\lower0.6ex\hbox{$\buildrel {\textstyle >}
 \over {\scriptstyle \sim}$}}}
\def\ls{\mathrel{\lower0.6ex\hbox{$\buildrel {\textstyle <}
 \over {\scriptstyle \sim}$}}}
\newcount\japequationnum
\global\japequationnum=0
\def\bookdisp#1$${\leftline{\hfill{$\displaystyle#1$}
    \global\advance\japequationnum by 1
    \hfill (\the\japequationnum )}$$}
\everydisplay{\bookdisp}
\def\japsub{\rm\scriptscriptstyle}

\def\hompc{{\,h\,\rm Mpc^{-1}}}

\def\japsub{\rm\scriptscriptstyle}

\newcommand{\plotter}[2]{\centering \leavevmode \epsfxsize=#2\textwidth \epsfbox{#1}\medskip}
\newcommand{\japplottwo}[2]{\centering \leavevmode 
\epsfxsize=0.49\textwidth \epsfbox{#1}
\hglue 1em
\epsfxsize=0.49\textwidth \epsfbox{#2}
\medskip}

\def\m@th{\mathsurround=0pt }
\def\eqalign#1{\null\,\vcenter{\openup1\jot \m@th
 \ialign{\strut\hfil$\displaystyle{##}$&$\displaystyle{{}##}$\hfil
 \crcr#1\crcr}}\,}
\def\cases#1{\left\{\,\vcenter{\normalbaselines\m@th
    \ialign{$##\hfil$&\quad##\hfil\crcr#1\crcr}}\right.}

\begin{document}

\title[LSS in the universe]{Large-scale structure and matter\\
in the universe}

\author[J.A. Peacock]{J.A. Peacock}

\affiliation{Institute for Astronomy, University of Edinburgh,\\
Royal Observatory, Edinburgh EH9 3HJ, UK}

\label{firstpage}

\maketitle

\begin{abstract}{Cosmology -- Galaxies: clustering}
This paper summarizes the physical mechanisms that 
encode the type and quantity of cosmological matter
in the properties of large-scale structure, and reviews 
the application of such tests to current datasets.
The key lengths of the horizon size at matter-radiation
equality and at last scattering determine the total
matter density and its ratio to the relativistic density;
acoustic oscillations can diagnose whether the matter is
collisionless, and small-scale structure or its absence can
limit the mass of any dark-matter relic particle.
The most stringent constraints come from combining data
on present-day galaxy clustering with data on CMB anisotropies.
Such an analysis breaks the degeneracies inherent in either
dataset alone, and proves that the universe is very close to
flat. The matter content is accurately consistent with
pure Cold Dark Matter, with about 25\% of the critical
density, and fluctuations that are scalar-only, adiabatic
and scale-invariant. It is demonstrated that these
conclusions cannot be evaded by adjusting either the
equation of state of the vacuum, or the total relativistic
density.
\end{abstract}

\sec{Dark matter and growth of structure}

\ssec{Gravitational instability and transfer function}

The simplest explanation for the large-scale structure in the
galaxy distribution is
gravitational instability acting on some small
initial departures from homogeneity. 
This model has now reached a stage of considerable 
sophistication, and accounts impressively for the
very detailed data emerging from the current generation of
surveys.

The ability to use observations of large-scale structure to
measure the matter content of the universe depends on
understanding the characteristic scales that should be
introduced by gravitational instability. These are most
clearly seen in Fourier space, where the fractional
density contrast is $\delta({\bf x}) = \sum \delta_k \exp(i{\bf k\cdot x})$,
with $\rho = \bar\rho\,(1+\delta)$. The power spectrum is
$P(k)\equiv|\delta_k|^2$, conveniently expressed in the
dimensionless form, where $\Delta^2(k)\propto k^3 P(k)$ is
the variance in $\delta$ per $\ln k$. The assumption is that
this can be decomposed into a primordial component of
power-law form, and a transfer function:
$$
\Delta^2(k) \propto k^{3+n} T^2(k),
$$
where the function $T(k)$ contains the information about
the matter content. A generic expectation for the primordial
fluctuations is that $n=1$, so that the power spectrum
of potential fluctuations is 3D flicker noise, and the
deviations of the metric from flatness are fractal-like.
An interesting feature of inflation, which makes an
important test of the theory, is that small deviations
from exact $n=1$ behaviour may be expected.

The transfer function depends on a number of characteristics
of the primordial perturbations, both qualitative and quantitative:

\japitem{(1)} Perturbation mode.  The simplest choice is that
the initial fluctuations were adiabatic:
i.e. photon densities and matter densities were
compressed equally. This is the prediction of single-field
inflation models. In more complex models, it is possible that
the radiation is left
unperturbed, and only the matter fluctuates.
Such isocurvature modes or
entropy perturbations
match the CMB anisotropy data poorly, so
we will neglect them. However, a small admixture of isocurvature
modes can always be tolerated, and this can widen the space
of allowed models 
(e.g. Bucher, Moodley \& Turok 2002).

\japitem{(2)} Relativistic content. It turns out that
most of the characteristics of cosmological perturbations
were set at high redshifts, when the relativistic particle
content (at least photons plus light neutrinos) was dynamically
important.

\japitem{(3)} Baryonic content.
At early times, the baryonic plasma is strongly coupled to the
photons via Thomson scattering, thus acting as a fluid with
a sound speed of up to $c/\sqrt{3}$. Acoustic oscillations in
this fluid leave measurable features in the transfer function.

\japitem{(4)} Collisionless content. Growth of perturbations in
weakly-interacting dark matter proceed in a simpler was, and they
do not support pressure-driven oscillations. Also, the key
free-streaming property of collisionless particles can lead to 
erasure of small-scale perturbations.

\japitem{(5)} Vacuum-energy content.
In general, a homogeneous background does not influence the scale
dependence of perturbation growth -- rather, the growth rate of
perturbations of all wavelengths is altered. Thus, the main
influence of the vacuum energy is via the overall amplitude
of fluctuations plus, in the case of the CMB, the conversion
from spatial scale of fluctuations observed at redshift 1100
to angle subtended today.

\ssec{Characteristic scales}

The transfer function for models with the full above list
of ingredients was first computed accurately by Bond \& Szalay (1983),
and is today routinely available via public-domain codes
such as {\sc cmbfast} (Seljak \& Zaldarriaga 1996).
Some illustrative results are shown in Figure~1. Leaving aside
the isocurvature models, all adiabatic cases have $T\rightarrow 1$
on large scales -- i.e. there is growth at the universal
rate (which is such that the amplitude of potential perturbations
is constant until the vacuum starts to be important at $z\ls 1$).
The different shapes of the functions can be understood intuitively
in terms of a few special length scales, as follows:

\begin{figure}[ht]
\plotter{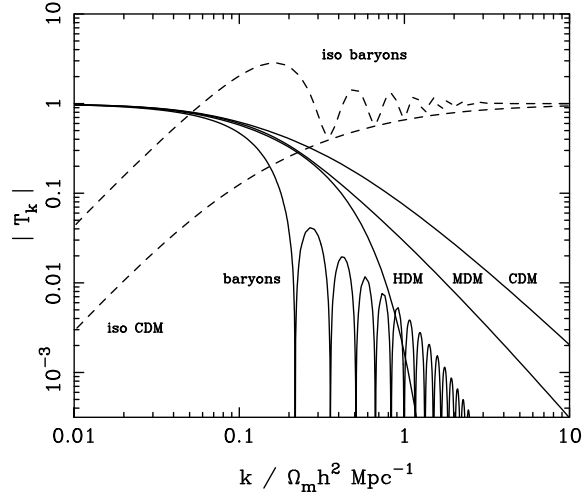}{0.6}
\caption{
A plot of transfer functions for various
adiabatic models, in which $T_k\rightarrow 1$ at small $k$.
A number of possible matter contents are illustrated:
pure baryons; pure CDM; pure HDM.
For dark-matter models, the characteristic wavenumber scales
proportional to $\Omega_m h^2$, marking the break 
scale corresponding to the horizon length at 
matter-radiation equality. The scaling for baryonic
models does not obey this exactly; the plotted case corresponds
to $\Omega_m=1$, $h=0.5$.
}
\end{figure}

{\bf (1) Horizon length at matter-radiation equality.}
The main bend visible in all 
transfer functions is due to the M\'esz\'aros effect, which 
arises because the universe is radiation
dominated at early times. 
Fluctuations in the matter can only grow if dark
matter and radiation fall together. This does not
happen for perturbations of small wavelength, because
photons and matter can separate.
Growth only occurs for perturbations of
wavelength larger than the horizon distance, where there has been no time for the
matter and radiation to separate.
The relative diminution in fluctuations at high $k$ is the amount of growth 
that is missed between horizon entry and $z_{\rm eq}$, and
this change is easily shown to be $\propto k^2$.
The approximate limits of the CDM
transfer function are therefore
$$
T_k \simeq \cases{
1 \quad\quad &\hbox{$kD_{\japsub  H}(z_{\rm eq})\ll 1$} \cr
[kD_{\japsub  H}(z_{\rm eq})]^{-2} \quad\quad &\hbox{$kD_{\japsub  H}(z_{\rm eq})\gg 1$}. \cr
}
$$

This process continues, until the universe becomes matter
dominated at $z_{\rm eq}=23\,900\,\Omega_m h^2$.
We therefore expect a characteristic `break' in the
fluctuation spectrum around the comoving horizon length at this time:
$$
D_{\japsub  H}(z_{\rm eq})  =
(\sqrt{2}-1)
\frac{2c}{H_0}(\Omega_m z_{\rm eq})^{-1/2} =
16\, (\Omega_m h^2)^{-1}{\rm Mpc}.
$$
Since distances in cosmology always scale as $h^{-1}$, this means
that $\Omega_m h$ should be observable.

{\bf (2) Free-streaming length.}
This relatively gentle filtering away of the initial
fluctuations is all that applies to a universe dominated
by Cold Dark Matter, in which random velocities are
negligible. 
A CDM universe thus contains fluctuations
in the dark matter on all scales, and
structure formation proceeds via 
hierarchical process in which nonlinear structures
grow via mergers.

Examples of CDM would be thermal relic WIMPs with masses
of order 100~GeV. Relic particles that were never in
equilibrium, such as axions, also come under
this heading, as do more exotic possibilities
such as primordial black holes.
A more interesting case arises when thermal relics
have lower masses.
For collisionless dark matter, perturbations can be
erased simply by free streaming: random
particle velocities cause blobs to disperse.
At early times ($kT>mc^2$), the particles will
travel at $c$, and so any perturbation that has
entered the horizon will be damped.
This process ceases when the particles
become non-relativistic, so that perturbations
are erased up to proper lengthscales of 
$ \simeq ct(kT = mc^2)$.
This translates to a comoving horizon scale
($2ct/a$ during the radiation era) at $kT = mc^2$ of
$$
L_{\rm free-stream} = 112\, (m/{\rm eV})^{-1}\, {\rm Mpc}
$$
(in detail, the appropriate figure for neutrinos will
be smaller by $(4/11)^{1/3}$ since they have a smaller
temperature than the photons).
A light neutrino-like relic that decouples while
it is relativistic satisfies
$$
\Omega_\nu h^2 = m / 93.5\,{\rm eV}
$$
Thus, the damping
scale for HDM (Hot Dark Matter) is of order the bend scale.
Alternatively, if the particle decouples sufficiently
early, its relative number density is boosted by annihilations,
so that the critical particle  mass 
to make $\Omega_m=1$ can be boosted to
around 1--10~keV (Warm Dark Matter).
The existence of galaxies at $z\simeq 6$ tells us that the
coherence scale must have been below about 100~kpc, so
WDM is close to being ruled out.
A similar constraint is obtained from small-scale
structure in the Lyman-alpha forest (Narayanan et al. 2000):
$m > 0.75$~keV.

A more interesting (and probably more practically relevant) case
is when the dark matter is a mixture of hot and cold
components. The free-streaming length for the hot
component can therefore be very large, but within
range of observations. The dispersal of HDM fluctuations
reduces the CDM growth rate on all scales below
$L_{\rm free-stream}$ -- or, relative to small
scales, there is an enhancement in large-scale power.

{\bf (3) Acoustic horizon length.}
The horizon at matter-radiation equality also enters in
the properties of the baryon component. Since the sound speed
is of order $c$, the largest scales that can undergo a single
acoustic oscillation are of order the horizon. 
The transfer function for a purely baryonic universe shows
large modulations, reflecting the number of oscillations
that were completed before the universe became
matter dominated and the pressure support dropped.
The lack of such large modulations in real data is
one of the most generic reasons for believing in 
collisionless dark matter. Acoustic oscillations
persist even when baryons are subdominant, however, and
can be detectable as lower-level modulations in the transfer
function (e.g. Goldberg \& Strauss 1998; Meiksin et al. 1999).

{\bf (4) Silk damping length.}
Acoustic oscillations are also damped on small scales, where
the process is called Silk damping: the
mean free path of photons due to scattering by the plasma
is non-zero, and so radiation can diffuse out of a
perturbation, convecting the plasma with it. This effect
can be seen in Figure 1 at $k\sim 10 k_{\japsub H}$.

\sec{Comparison with 2dFGRS data}

\ssec{Survey overview}

The largest dataset for which a thorough comparison with the
above picture has been made is the
2dF Galaxy Redshift Survey (2dFGRS).
This survey was designed around the 2dF multi-fibre spectrograph on the
Anglo-Australian Telescope, which is capable of observing up to 400
objects simultaneously over a 2~degree diameter field of view. 
For details of
the instrument and its performance 
see {\tt http://www.aao.gov.au/2df/}, and also
Lewis et~al.\ (2002).
The source catalogue for the survey is a revised and extended version of
the APM galaxy catalogue (Maddox et~al.\ 1990a,b,c); this
includes over 5~million galaxies down to $b_{\japsub J}=20.5$ in both
north and south Galactic hemispheres over a region of almost
$10^4\, {\rm deg}^2$.
The $b_{\japsub J}$
magnitude system 
is related to the Johnson--Cousins system by $b_{\japsub J} = B -0.304(B-V)$,
where the colour term is estimated from comparison with the SDSS Early
Data Release (Stoughton et al. 2002).

\begin{figure}[ht]
\plotter{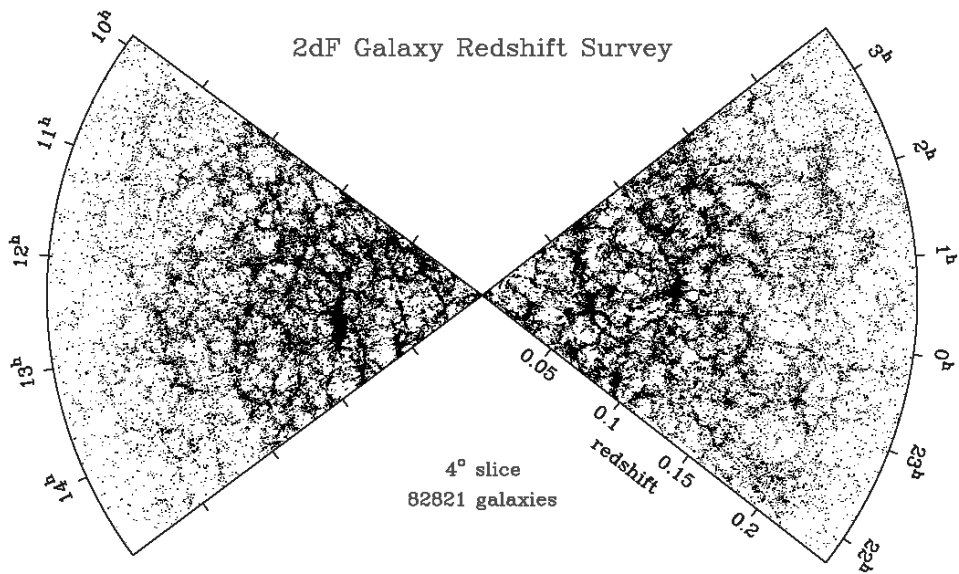}{1.0}
\caption{The distribution of galaxies in part of the 2dFGRS:
slices $4^\circ$ thick, centred at declination
$-2.5^\circ$ in the NGP and $-27.5^\circ$ in the SGP.
This magnificently detailed image of large-scale structure
provides the basis for measuring the shape of the
primordial fluctuation spectrum and hence constraining the
matter content of the universe.}
\end{figure}

The 2dFGRS geometry consists of two contiguous
declination strips, plus 100 random 2-degree fields. One strip is in the
southern Galactic hemisphere and covers approximately
75$^\circ$$\times$15$^\circ$ centred close to the SGP at
($\alpha, \delta$)=($01^h$,$-30^\circ$); the other strip is in the northern
Galactic hemisphere and covers $75^\circ \times 7.5^\circ$ centred at
($\alpha, \delta$)=($12.5^h$,$+0^\circ$). The 100 random fields are spread
uniformly over the 7000~deg$^2$ region of the APM catalogue in the
southern Galactic hemisphere. 
The sample is limited to be brighter than an extinction-corrected
magnitude of $b_{\japsub J}=19.45$ (using the extinction maps of Schlegel et~al.\
1998). This limit gives a good match between the density on the sky of
galaxies and 2dF fibres.

After an extensive period of commissioning of the 2dF instrument,
2dFGRS observing began in earnest in May 1997, and terminated
in April 2002. 
In total, observations were made of 899 fields,
yielding redshifts and identifications for 232,529 galaxies, 13976 stars
and 172 QSOs, at an overall completeness of 93\%. 
The galaxy redshifts are assigned a quality flag from 1 to 5,
where the probability of error is highest at low $Q$. Most analyses
are restricted to $Q\ge 3$ galaxies, of which there are currently
221,496.
An interim data release took place in July 2001,
consisting of approximately 100,000 galaxies (see Colless et al. 2001
for details). A public release of the full photometric and spectroscopic
database is scheduled for July 2003.
The completed 2dFGRS yields a striking
view of the galaxy distribution over large cosmological volumes.
This is illustrated in
Figure~2, which shows the projection of a subset of
the galaxies in the northern and southern strips onto $(\alpha,z)$
slices. This picture is the culmination of decades of effort in 
the investigation of large-scale structure, and we are
fortunate to have this detailed view for the first time.

\begin{figure}[ht]
\plotter{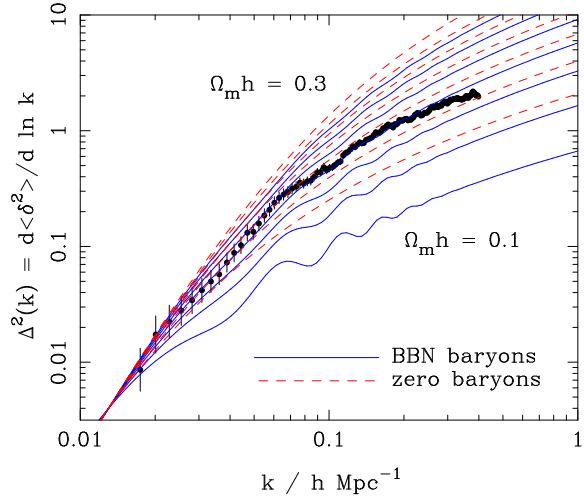}{0.6}
\caption{The 2dFGRS redshift-space dimensionless power spectrum, 
$\Delta^2(k)$,
estimated according to the FKP procedure. The solid points
with error bars show the power estimate. The window
function correlates the results at different $k$ values,
and also distorts the large-scale shape of the power spectrum
An approximate correction for the latter effect has been applied.
The solid and dashed lines show various CDM models, all assuming
$n=1$. For the case with non-negligible baryon content,
a big-bang nucleosynthesis value of $\Omega_b h^2=0.02$ is
assumed, together with $h=0.7$. A good fit is clearly obtained
for $\Omega_m h \simeq 0.2$. Note that the observed power at
large $k$ will be boosted by nonlinear effects, but damped by 
small-scale random peculiar velocities. It appears that these
two effects very nearly cancel, but model fitting is generally
performed only at $k<0.15 \hompc$ in order to avoid these complications.}
\end{figure}

\ssec{The 2dFGRS power spectrum}

Perhaps the key aim of the 2dFGRS was to perform an accurate
measurement of the 3D clustering power spectrum, in order
to improve on the APM result,
which was deduced by deprojection of angular
clustering (Baugh \& Efstathiou 1993, 1994). 
The results of this direct estimation of the 3D power
spectrum are shown in Figure~3 (Percival et al. 2001).
This power-spectrum estimate uses the FFT-based approach
of Feldman, Kaiser \& Peacock (1994; FKP), and needs to be interpreted
with care. Firstly, it is a raw redshift-space estimate, so
that the power beyond $k\simeq 0.2 \hompc$ is severely damped
by smearing due to peculiar velocities, as well as being
affected by nonlinear evolution.
Finally, the FKP estimator yields the
true power convolved with the window function. This
modifies the power significantly at large scales (roughly
a 20\% correction). An approximate correction for
this has been made in Figure~3. 

\begin{figure}[ht]
\japplottwo{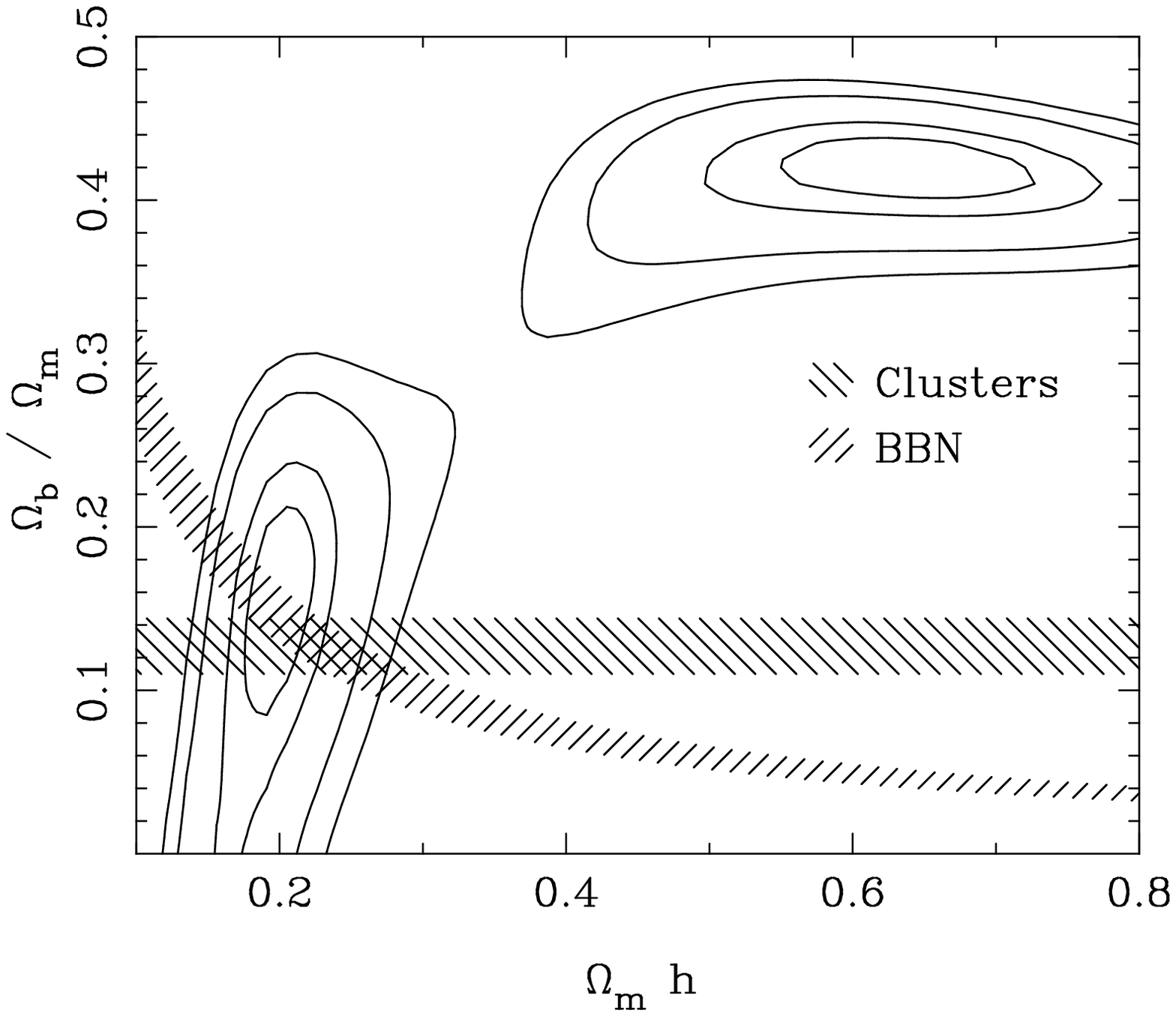}{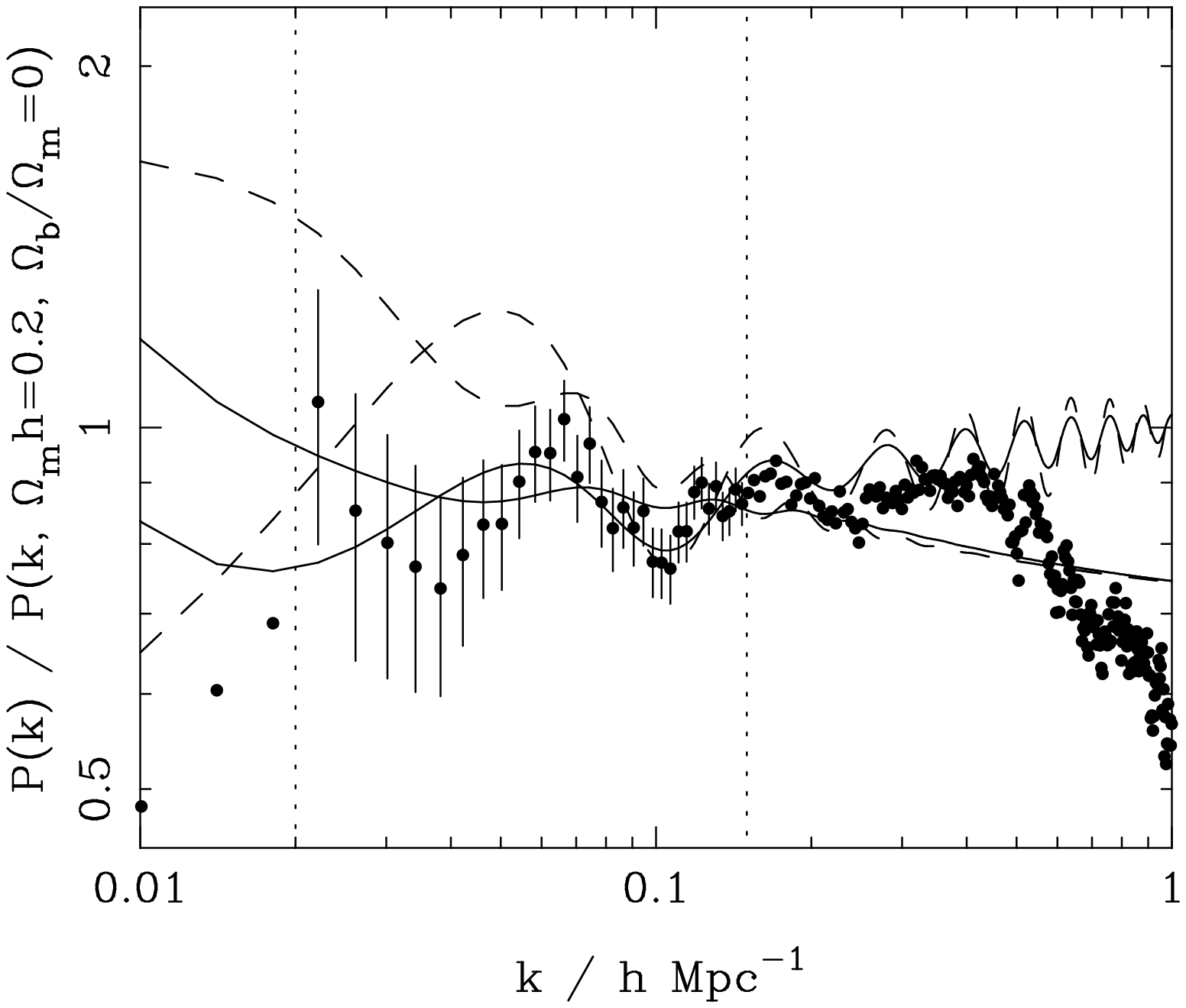}
\caption{Likelihood contours for the best-fit linear
  CDM fit to the 2dFGRS power spectrum
  over the region $0.02<k<0.15$. Contours are plotted
  at the usual positions for one-parameter confidence of 68\%, and
  two-parameter confidence of 68\%, 95\% and 99\% (i.e. $-2\ln({\cal
  L}/{\cal L_{\rm max}}) = 1, 2.3, 6.0, 9.2$). We have marginalized
  over the missing free parameters ($h$ and the power spectrum
  amplitude).
  A prior on $h$ of $h=0.7\pm 10\%$ was assumed. 
  This result is compared to estimates from X-ray cluster
  analysis (Evrard 1997) and big-bang nucleosynthesis (Burles et al.  2001).
The second panel shows the 2dFGRS data compared with the two preferred models from
  the Maximum Likelihood fits convolved with the window function
  (solid lines). The unconvolved models are also shown (dashed
  lines). The $\Omega_m h \simeq 0.6$, $\Omega_b/\Omega_m=0.42$,
  $h=0.7$ model has the higher bump at $k\simeq 0.05\hompc$. The
  smoother $\Omega_m h \simeq 0.20$, $\Omega_b/\Omega_m=0.15$, $h=0.7$
  model is a better fit to the data because of the overall shape.
A preliminary analysis of the complete final 2dFGRS sample yields
a slightly smoother spectrum than the results shown
here (from Percival et al. 2001), so that the high-baryon solution becomes
disfavoured.
}
\end{figure}

\ssec{CDM model fitting} 

The fundamental assumption is that, on large scales, linear biasing
applies, so that the nonlinear galaxy power spectrum in redshift space has a shape
identical to that of linear theory in real space.
This assumption is valid for $k<0.15\hompc$;
the detailed justification comes from analyzing realistic 
mock data derived from $N$-body simulations (Cole et al. 1998).
The free parameters in fitting CDM models are thus the primordial spectral
index, $n$, the Hubble parameter, $h$, the total matter
density, $\Omega_m$, and the baryon fraction, $\Omega_b/\Omega_m$.
Note that the vacuum energy does not affect the constraints. Initially, we
show results assuming $n=1$; this assumption is relaxed later.

An accurate model comparison requires
the full covariance matrix of the data, because
the convolving effect of the window function 
causes the power at adjacent $k$ values to be correlated.
This covariance matrix was estimated by applying the survey window to a
library of Gaussian realisations of linear density fields, and
checked against a set of mock catalogues.
It is now possible to explore the space of CDM models, and
likelihood contours in $\Omega_b/\Omega_m$ versus $\Omega_mh$
are shown in Figure~4. At each point in this
surface we have marginalized by integrating the likelihood surface
over the two free parameters, $h$ and the power spectrum
amplitude. 
We have added a Gaussian prior $h=0.7\pm
10\%$, representing external constraints such as the HST key project
(Freedman et al. 2001); this has only a minor effect on the results.

Figure~4a shows that there is a degeneracy between
$\Omega_mh$ and the baryonic fraction $\Omega_b/\Omega_m$. However, there
are two local maxima in the likelihood, one with $\Omega_mh \simeq 0.2$
and $\sim 20\%$ baryons, plus a secondary solution $\Omega_mh \simeq 0.6$
and $\sim 40\%$ baryons. The high-density model can be rejected through a variety
of arguments, and the preferred solution is
$$
  \Omega_m h = 0.20 \pm 0.03; \quad\quad \Omega_b/\Omega_m = 0.15 \pm 0.07.
$$
The 2dFGRS data are compared to the best-fit linear power spectra
convolved with the window function in Figure~4b. The low-density
model fits the overall shape of the spectrum with relatively small
`wiggles', while the solution at $\Omega_m h \simeq 0.6$ provides a
better fit to the bump at $k\simeq 0.065\hompc$, but fits the overall
shape less well.
A preliminary analysis of $P(k)$ from the full final dataset
shows that $P(k)$ becomes smoother: 
the high-baryon solution becomes disfavoured, and
the uncertainties narrow slightly around the lower-density solution:
$\Omega_m h = 0.18 \pm 0.02$; $\Omega_b/\Omega_m = 0.17 \pm 0.06$.

It is interesting to compare these conclusions with other
constraints. These are shown on Figure~4, assuming 
$h=0.7\pm 10\%$.
Latest estimates of the Deuterium to Hydrogen ratio in QSO spectra
combined with big-bang nucleosynthesis theory predict $\Omega_bh^2 =
0.020\pm 0.001$ (Burles et al. 2001), which translates to the
shown locus of $f_{\japsub B}$ vs $\Omega_m h$. X-ray
cluster analysis predicts a baryon fraction
$\Omega_b/\Omega_m=0.127\pm0.017$ (Evrard 1997) which is within
$1\sigma$ of our value. These loci intersect very close
to our preferred model.

Perhaps the main point to emphasise here is that the 2dFGRS results are not
greatly sensitive to the assumed tilt of the primordial spectrum. 
As discussed below, CMB data show that $n=1$ is a very
good approximation; in any case, very substantial tilts ($n\simeq 0.8$)
are required to alter the conclusions significantly.

\ssec{Robustness of results}

The main residual worry about accepting the above conclusions is 
probably whether the assumption of linear bias can really be valid. 
In general, concentration towards higher-density regions both
raises the amplitude of clustering, but also steepens the correlations,
so that bias is largest on small scales. 
A simple model that illustrates this is to assume that the density
field is a lognormal process. A nonlinear transformation 
$\rho \rightarrow \rho^b$ then gives a correlation function
$1+\xi \rightarrow (1+\xi)^{b^2}$ (Mann, Peacock \& Heavens 1998).
We need to be clear of the regime in which the bias depends
on scale.

One way in which
this issue can be studied is to consider subsamples with very
different degrees of bias. Colour information has recently 
been added to the 2dFGRS database using SuperCosmos scans of the
UKST red plates (Hambly et al. 2001), and a division at rest-frame photographic
$B-R=0.85$ nicely separates ellipticals from spirals.
Figure~5 shows the power spectra for the 2dFGRS divided in this
way. The shapes are almost identical (perhaps not so surprising,
since the cosmic variance effects are closely correlated in these
co-spatial samples). However, what is impressive is that the
relative bias is almost precisely independent of scale,
even though the red subset is rather strongly biased
relative to the blue subset (relative $b\simeq 1.4$). This provides some
reassurance that the large-scale $P(k)$ reflects the underlying
properties of the dark matter, 
rather than depending on the particular class
of galaxies used to measure it.

\begin{figure}[ht]
\japplottwo{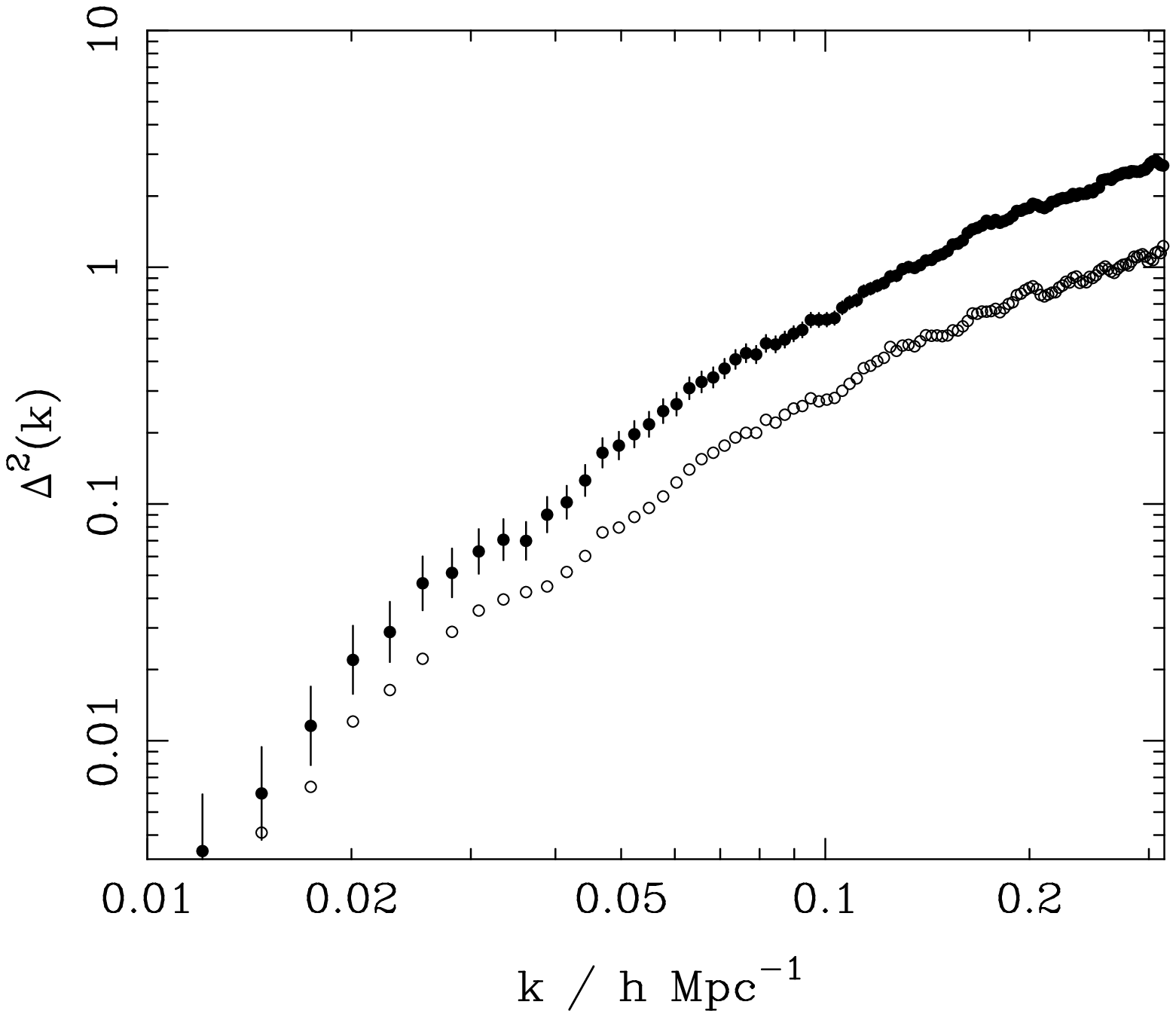}{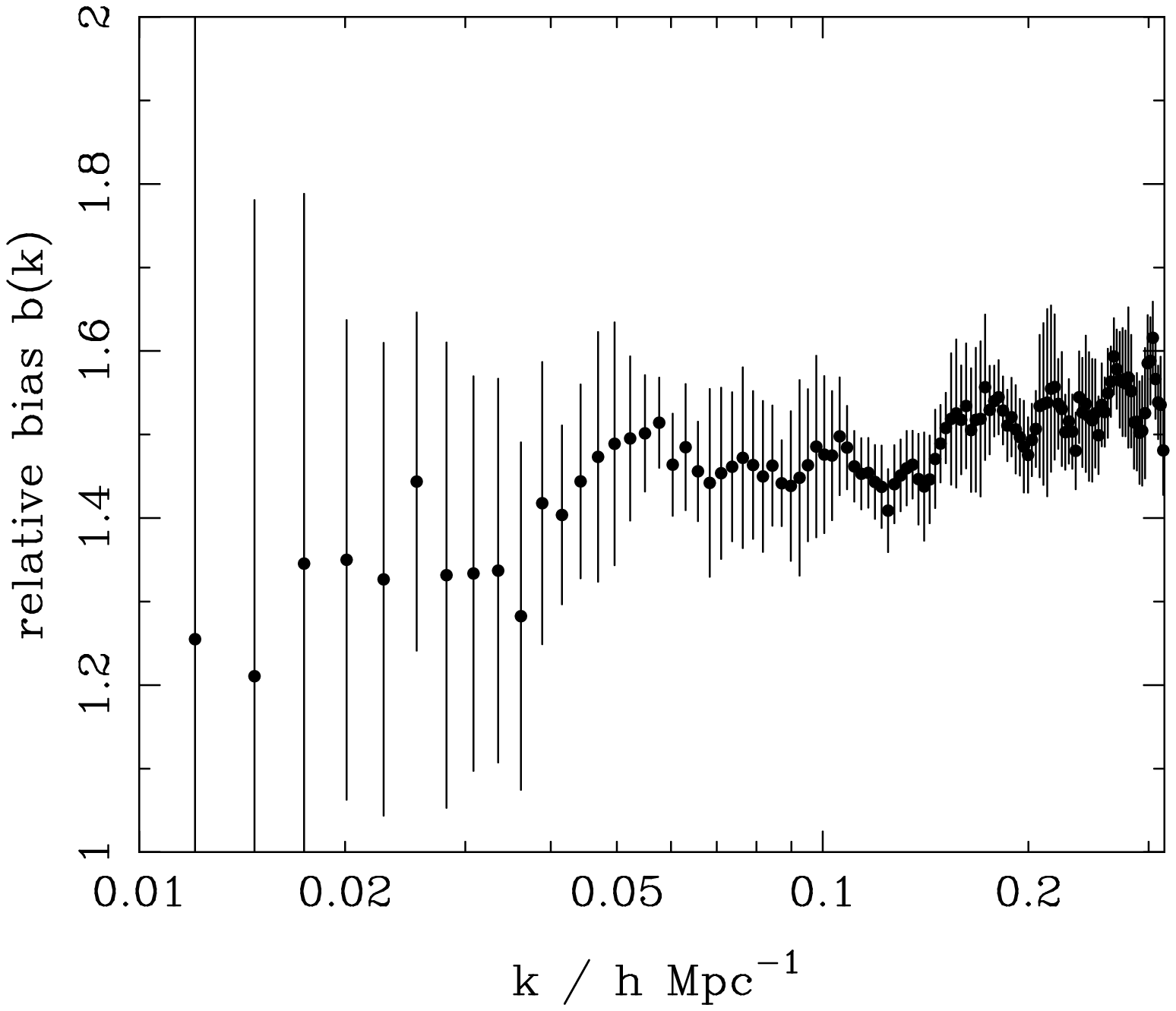}
\caption{
The power spectra of red galaxies (filled circles) and blue galaxies (open circles),
divided at photographic $B-R=0.85$.
The shapes are strikingly similar, and
the square root of the ratio yields the right-hand panel:
the relative bias in redshift space of red and blue galaxies.
The error bars are obtained by a jack-knife analysis. The relative
bias is consistent with a constant value of 1.4 over the range used for
fitting of the power-spectrum data ($0.015 < k < 0.15 \hompc$).
}
\end{figure}

\sec{Combination with the CMB and cosmological parameters}

\ssec{Parameter degeneracies}

The 2dFGRS power spectrum contains important information about the
key parameters of the cosmological model, but we have seen that additional
assumptions are needed, in particular the values of $n$ and $h$.
Observations of CMB
anisotropies can in principle measure 
most of the cosmological parameters, and
combination with the 2dFGRS can lift most of the degeneracies inherent
in the CMB-only analysis. It is therefore of interest to see
what emerges from a joint analysis.
These issues are discussed in Efstathiou et al. (2002).
The CMB data alone contain two important degeneracies:
the `geometrical' and `tensor' degeneracies.

{\bf Geometrical degeneracy}
In the former case, one can evade the
commonly-stated CMB conclusion that the
universe is flat, by adjusting both $\Lambda$ and $h$ to
extreme values (Zaldarriaga et~al. 1997; Bond et~al. 1997; Efstathiou \& Bond 1999).  
The normal flatness argument takes the
comoving horizon size at last scattering
$$
D_{\japsub LS} = \frac{2c}{\Omega_m^{1/2} H_0}\,(1+z_{\japsub LS})^{-1/2} \simeq
184(\Omega_m h^2)^{-1/2}\; {\rm Mpc}
$$
and divides it by the present-day horizon size for a zero-$\Lambda$ universe,
$$
D_{\japsub H} = \frac{2c}{\Omega_m H_0},
$$
to yield a main characteristic angle that scales as $\Omega_m^{1/2}$.
Large curvature (i.e. low $\Omega_m$) is ruled out because the
main peak in the CMB power spectrum is not seen at very small
angles. However, introducing vacuum energy changes the conclusion.
If we take a
family of models with fixed initial perturbation spectra, fixed
physical densities $\omega_m \equiv \Omega_m h^2$,  
$\omega_b \equiv \Omega_b h^2$,
and vary both $\Omega_v$ and the curvature to keep a fixed value of
the angular size distance to last scattering, then the
resulting CMB power spectra are identical (except for the Integrated
Sachs-Wolfe effect at very low multipoles, 
and second-order effects at high $\ell$).
This degeneracy occurs because the physical densities
control the structure of the perturbations in physical Mpc at last
scattering, while curvature, $\Omega_v$ and $\Omega_m$
govern the proportionality
between length at last scattering and observed angle.
In order to break the degeneracy, additional information is
needed. This could be in the form of external data on the Hubble
constant, but the most elegant approach is to add the 2dFGRS data,
so that conclusions are based only on the shapes of power spectra.
Efstathiou et al. (2002) show that doing this
yields a total density ($|\Omega-1|<0.05$) at 95\% confidence.
We can therefore be confident that the universe is very nearly flat;
hereafter it will be assumed that this is exactly true.

{\bf Tensor degeneracy}
The next most critical question for the CMB is whether the
temperature fluctuations are scalar-mode only, or whether
there could be a significant tensor signal. The tensor modes
lack acoustic peaks, so they reduce the relative amplitude
of the main peak at $\ell=220$.
A model with a large
tensor component can however be made to resemble a zero-tensor model
by applying a large blue tilt ($n>1$) and a high baryon content.
Efstathiou et al. (2002) show that adding the 2dFGRS
data weakens this degeneracy, but does not completely remove it.
This is reasonable, since the 2dFGRS data alone constrain
the baryon content weakly.

The importance of tensors
will of course be one of the key questions for cosmology over the
next several years, but it is interesting to consider the limit
in which these are negligible. In this case, the standard model
for structure formation contains a vector of only 6 parameters:
${\bf p} = (n_s, \Omega_m, \Omega_b, h, Q, \tau)$.
Of these, the optical depth to last scattering, $\tau$, is almost entirely
degenerate with the normalization, $Q$.
The remaining four parameters are pinned down very precisely:
using a compilation of pre-WMAP CMB data plus the 2dFRGS power spectrum,
Percival et al. (2002) obtained
$$
(n_s, \Omega_c, \Omega_b, h) =
(0.963\pm 0.042, 0.115\pm 0.009, 0.021\pm0.002, 0.665 \pm 0.047),
$$
or an overall density parameter of $\Omega_m=0.313 \pm 0.055$.

It is remarkable how well these figures agree with completely
independent determinations: $h=0.72\pm 0.08$ from the HST key project
(Mould et al. 2000; Freedman et al. 2001);
$\Omega_b h^2 =0.020 \pm 0.001$ (Burles et al. 2001).
This gives confidence that the tensor component must
indeed be sub-dominant.
For further details of this analysis, see Percival et al. (2002).

\ssec{The horizon angle degeneracy}

For flat models, there is a degeneracy that is related (but not identical) to the geometrical
degeneracy, being very closely related to the location of the acoustic peaks.
The angular scale of these peaks
depends on the ratio between the horizon size at last scattering
and the present-day horizon size for flat models:
$$
\theta_{\japsub H} = 
D_{\japsub H}(z_{\japsub LS})/D_{\japsub H}(z=0); \quad\quad
D_{\japsub H}(z=0) = \frac{2c}{H_0}\Omega_m^{-0.4}.
$$
(using the approximation of Vittorio \& Silk 1985).
This yields an angle scaling as $\Omega_m^{-0.1}$, so that the
scale of the acoustic peaks is apparently almost independent
of the main parameters.

However, this argument is not complete because the earlier
expression for $D_{\japsub H}(z_{\japsub LS})$ assumes that
the universe is completely matter dominated at last scattering.
The comoving sound horizon size at last scattering is defined by
(e.g. Hu \& Sugiyama 1995)
$$
 D_{\japsub S}(z_{\japsub LS}) \equiv \frac{1}{H_0 \Omega_m^{1/2}}
   \int_0^{a_{\japsub LS}} \frac{c_{\japsub S}}{(a + a_{\rm eq})^{1/2} } \, da
$$
where vacuum energy is neglected at these high redshifts;
the expansion factor $a \equiv (1+z)^{-1}$ and
$a_{\japsub LS}, a_{\rm eq}$ are the values at last scattering and
matter-radiation equality respectively.
In practice, $z_{\japsub LS}\simeq 1100$ independent of the matter
and baryon densities, and $c_{\japsub S}$ is fixed by 
$\Omega_b$. Thus the main effect is that $a_{\rm eq}$ depends
on $\Omega_m$.
Dividing by $D_{\japsub H}(z=0)$ therefore
gives the angle subtended today by the light horizon as
$$
  \theta_{\japsub H} \simeq  \frac{\Omega_m^{-0.1}}{\sqrt{1+z_{\japsub LS}}}
     \left[\sqrt{1 + \frac{a_{\rm eq}}{a_{\japsub LS}} } -
  \sqrt{\frac{a_{\rm eq}}{a_{\japsub LS}} }\, \right],
$$
where  $z_{\japsub LS} = 1100$ and $a_{\rm eq} = (23900 \,\omega_m)^{-1}$.
This remarkably simple result captures  most
of the parameter dependence of CMB peak locations within
flat $\Lambda$CDM  models.
Differentiating this equation near a fiducial $\omega_m = 0.147$ gives
$$
  \left.\frac{\partial\ln\theta_{\japsub
  H}}{\partial\ln\Omega_m}\right|_{\omega_m}
   = -0.1;
\quad
\quad
  \left.\frac{\partial\ln\theta_{\japsub
  H}}{\partial\ln\omega_m}\right|_{\Omega_m}=
  \frac{1}{2} \left( 1 + \frac{a_{\japsub LS}}{a_{\rm eq}}\right)^{-1/2} =  +0.24 ,
$$
in good agreement with the numerical derivatives 
in Eq.~(A15) of Hu et~al. (2001).  

Thus for moderate variations from a `fiducial' model, the CMB peak
multipole number scales approximately as $\ell_{\rm peak} \propto \Omega_m^{-0.14}
h^{-0.48}$, i.e.  the condition for constant CMB peak location is well
approximated as 
$$
\Omega_m h^{3.4} = {\rm constant}.
$$
However, information about the
peak heights does alter this degeneracy slightly; the relative peak
heights are preserved at constant $\Omega_m$, hence the actual likelihood
ridge is a `compromise' between constant peak location (constant
$\Omega_m h^{3.4}$) and constant relative heights (constant $\Omega_m
h^2$); the peak locations have more weight in this compromise, leading
to a likelihood ridge along approximately $\Omega_m h^{3.0} \simeq {\rm const}$
(Percival et al. 2002).
It is now clear how LSS data combines with the CMB: $\Omega_m h^{3.4}$
is measured to very high accuracy already, and
Percival et al. deduced $\Omega_m h^{3.4}= 0.078$ with an error of
about 6\% using pre-WMAP CMB data. The first-year WMAP results
in fact prefer $\Omega_m h^{3.4}= 0.084$ (Spergel et al. 2003); the slight
increase arises because WMAP indicates that previous datasets around
the peak were on average calibrated low. 

In any case, the dominant error in $\Omega_m$ and $h$ depends
on what one chooses to add to the $\Omega_m h^{3.4}$ figure. 
The best approach given current knowledge is probably to combine
the WMAP $\Omega_m h^{3.4}= 0.084$ with the updated 2dFGRS
$\Omega_m h = 0.18 \pm 0.02$: this yields
$\Omega_m = 0.25 \pm 15\%$ and $h=0.73 \pm 5\%$.

\ssec{Matter fluctuation amplitude and bias}

The above conclusions were obtained by considering the shapes
of the CMB and galaxy power spectra. However, it is also of
great interest to consider the amplitude of mass fluctuations,
since a comparison with the galaxy power spectrum
allows us to infer the degree of bias directly.
This analysis was performed by Lahav et al. (2002).
Given assumed values for the cosmological parameters, the
present-day linear normalization of the mass spectrum (e.g. $\sigma_8$)
can be inferred.
It is convenient to define a corresponding measure
for the galaxies, $\sigma_{8{\rm g}}$, such that we can express the bias parameter
as
$$
b = \frac{\sigma_{8{\rm g}}}{\sigma_{8{\rm m}} }.
$$
In practice, we define $\sigma_{8{\rm g}}$ to be the value  required
to fit a CDM model to the power-spectrum data on linear scales ($0.02<k<0.15 \hompc$).
The amplitude of 2dFGRS galaxies in real space estimated
by Lahav et al. (2002) is
$\smash{\sigma_{8{\rm g}}^R} (L^*) = 0.76$,
with a negligibly small random error.
This assumes no evolution in $\smash{\sigma_{8{\rm g}}}$,
plus the luminosity dependence of clustering measured by 
Norberg et al. (2001).

The value of $\sigma_8$ for the dark matter can be deduced from the
CMB fits. Percival et al. (2002) obtain 
$$
 \sigma_8 \exp(- \tau) = 0.72 \pm 0.04,
$$
where the quoted error includes both data errors and theory
uncertainty. The WMAP value here is almost identical:
$\sigma_8 \exp(- \tau) = 0.71$, but no error is quoted
(Spergel et al. 2003).
The unsatisfactory feature is the degeneracy
with the optical depth to last scattering. For reionization
at $z=8$, we would have $\tau\simeq 0.05$;
it is not expected theoretically that $\tau$ can be hugely larger,
and popular models would place reionization between $z=10$ and 
$z=15$, or $\tau\simeq 0.1$ (e.g. Loeb \& Barkana 2001). 
One of the many impressive aspects of the WMAP results is
that they are able to infer $\tau=0.17\pm 0.04$ from large-scale
polarization. Taken at face value, $\tau=0.17$ would  argue for reionization
at $z=20$, but the error means that more conventional figures are far from
being ruled out.
Taking all this together, it seems reasonable to assume that
the true value of $\sigma_8$ is within a few \% of 0.80.
Given the 2dFGRS figure of $\smash{\sigma_{8{\rm g}}^R} = 0.76$,
this implies that $L^*$ galaxies are very nearly exactly unbiased.
Since there are substantial variations
in the clustering amplitude with galaxy type, this outcome
must be something of a coincidence.

Finally, this conclusion of near-unity bias was reinforced
in a completely independent way, by using the
measurements of the bispectrum of galaxies in the 2dFGRS
(Verde et al. 2002). As it is based on three-point
correlations, this statistic is sensitive to the filamentary
nature of the galaxy distribution -- which is a signature of
nonlinear evolution. One can therefore split the degeneracy between
the amplitude of dark-matter fluctuations and the
amount of bias.

\sec{Less-standard ingredients}

\ssec{Limits to the neutrino mass}

Even though a CDM-dominated universe matches the
data very well, there are many plausible variations
to consider. Probably the most interesting is the
neutrino mass: experimental data mean that at least one
neutrino must have a mass of $\gs 0.05$~eV, so that
$\Omega_\nu \gs 10^{-3}$ -- the same order of magnitude
as stellar mass.

\begin{figure}[ht]
\plotter{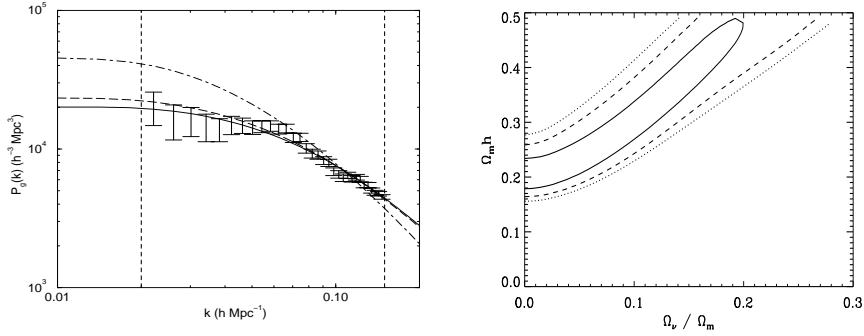}{0.95}
\caption{Results from Elgaroy et al. (2002),
who considered constraints on the neutrino mass from 2dFGRS.
The first panel shows 
Power spectra for $\Omega_\nu = 0$ (solid line),
$\Omega_\nu=0.01$ (dashed line), and $\Omega_\nu=0.05$ (dot-dashed line)
with amplitudes fitted to the 2dFGRS power spectrum data (vertical bars).
Other parameters are fixed at
$\Omega_{\rm m}=0.3$, $\Omega_\Lambda=0.7$, $h=0.7$, $\Omega_{\rm b}h^2=0.02$.
The vertical dashed lines limit the range in $k$ used in the fits.
The second panel shows
68\% (solid line), 95\% (dashed line) and 99\% (dotted line)
confidence contours in the plane of
$f_\nu\equiv \Omega_{\nu}/\Omega_{\rm m}$
and $\Gamma\equiv\Omega_{\rm m} h$,
with marginalization over $h$ and $\Omega_{\rm b}h^2$ using Gaussian
priors.
}
\end{figure}

As explained in Section 1, a non-zero neutrino mass can
lead to relatively enhanced large-scale power, beyond the
neutrino free-streaming scale. This is illustrated in 
Figure 6, taken from Elgaroy et al. (2002).
Broadly speaking, allowing a significant neutrino mass
changes the spectrum in a way that resembles lower density,
so there is a near-degeneracy between neutrino mass fraction
and $\Omega_m h$ (Figure~6b). A limit on the neutrino
fraction thus requires a prior on $\Omega_m h$. Based
on the cluster baryon fraction plus BBN, 
Elgaroy et al. adopt $\Omega_m<0.5$; together with the HST
Hubble constant, this yields a marginalized 95\% limit
of $f_\nu < 0.13$, or $m_\nu < 1.8$~eV.
Note that this is the sum of the eigenvalues of the mass matrix:
given neutrino oscillation results, the only way a cosmologically
significant density can arise is via a nearly degenerate
hierarchy, so this allows us to deduce $m_\nu < 0.6$~eV
for any one species.

\ssec{The equation of state of the vacuum}

So far, we have assumed that the vacuum energy is exactly
a classical $\Lambda$, or at any rate indistinguishable from
one. This is a highly reasonable prior: there is no reason
for the asymptotic value of any potential to go exactly
to zero, so one always needs to solve the classical cosmological
constant problem -- for which probably the only reasonable
explanation is an anthropic one (e.g. Vilenkin 2001).
Therefore, dynamical provision of $w\equiv p_v/\rho_v \ne -1$
is not needed. Nevertheless, one can readily take an
empirical approach to $w$ (treated as a constant for 
a frst approach).

\begin{figure}[ht]
\plotter{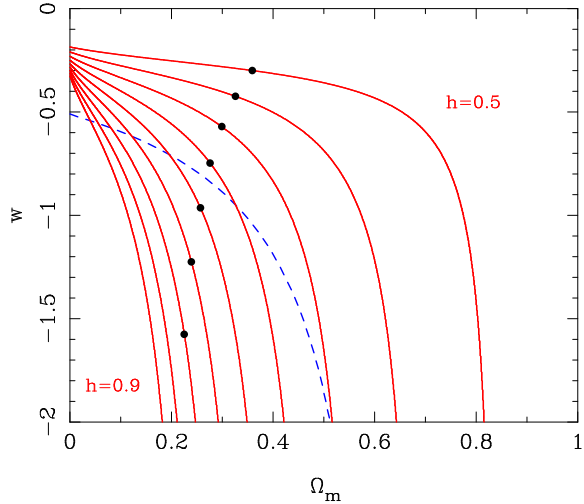}{0.6}
\caption{The $\Omega_m h^{3.4}$ degeneracy for flat models
gives an almost exact value of $\Omega_m$ from the CMB
is $h$ is known, assuming the vacuum to be effectively
a classical $\Lambda$ ($w=-1$). If $w$ is allowed to vary,
this becomes a locus on the $(\Omega_m, w)$ plane (similar
to the locus for best-fitting  flat models from the SNe, showed dotted).
Solid circles show values of $\Omega_m h$ that satisfy the
updated 2dFGRS constraint of 0.18 (suppressing error bars).}
\end{figure}

Figure~7 shows a simplified approach to this, plotting the
locus on $(w,\Omega_m)$ space that is required for
a given value of $h$ if the location of the main CMB acoustic peak
is known exactly. For $h\simeq 0.7$, this is very similar
to the locus derived from the SN Hubble diagram
(Garnavich et al. 1998).
The solid circles show the updated 2dFGRS constraint of
$\Omega_m h =0.18$. In order to match the data with
$w$ closer to zero, $\Omega_m$ must increase and $h$ must
decrease. The latter trend means that the HST Hubble constant
sets an upper limit to $w$ of about $-0.54$ (Percival et al. 2002).
This is very similar to the SNe constraint of Garnavich et al. (1998),
so the combined limit is already close to $w<-0.8$. The vacuum
energy is indeed looking rather similar to $\Lambda$.

\ssec{The total relativistic density}

Finally, an interesting aspect of Figure~7 is that it reminds
us of history. When the COBE detection was announced in 1992,
a popular model was `standard' CDM with $\Omega_m=1$, $h=0.5$.
As we see, this comes close to fitting the CMB data, and such
a model is not unattractive in some ways. Can we be sure it
is ruled out? Leaving aside the SNe data, one might think to
evade the 2dFGRS constraint by altering the total relativistic
content of the universe (for example, by the decay of a
heavy neutrino after nucleosynthesis). Since 2dFGRS measures
the horizon at matter-radiation equality, this will be changed.
If the radiation density is arbitrarily boosted by a factor
$X$, the constraint from LSS becomes
$$
(\Omega_m h)_{\rm apparent} = X^{-1/2} (\Omega_m h)_{\rm true}.
$$
Therefore $X\simeq 8$ is required to allow an Einstein--de Sitter
universe.

However, this argument fails, because it does not take into account
the effect of the extra radiation on the CMB. As argued above, the
location of the acoustic peaks depends on $a_{\rm eq}$, which depends
on $\omega_m$. However, if we change the radiation content, then
what matters is $\omega_m/X$. Thus, the CMB peak constraint now reads
$$
\Omega_m^{-0.1} (\omega_m/X)^{0.24} = {\rm constant};
$$
when combining LSS and CMB, everything is as before except that the
effective Hubble parameter is $h/X^{1/2}$. Thus, a model
with $\Omega_m=1$ but boosted radiation would only fit the CMB with
$h\simeq 0.5\sqrt{8} \simeq 1.4$, and the attractiveness
of a low age is lost. In any case, combining LSS and CMB would
give the same $\Omega_m \simeq 0.3$ independent of $X$, so it is
impossible to save models with $\Omega_m=1$ by this route.

Finally, it is interesting to invert this argument.
Since Percival et al. (2002) obtain an effective
$h$ of $0.665\pm 0.047$ and Freedman et al. (2001) measure
$h=0.72\pm 0.08$, we deduce
$$
1.68 X = 1.82 \pm 0.24.
$$
This convincingly rules out the $1.68 X=1$ that would apply
if the universe contained only photons, and amounts to
a detection of the neutrino background. In
terms of the number of neutrino species, this is
$N_\nu = 3.6 \pm 1.1$. A more precise result is of
course obtained from primordial nucleosynthesis, but this
applies at a much later epoch, thus constraining models
with decaying particles.

\sec{Conclusions}

The beautiful data on the large-scale structure of the universe
revealed in particular by the 2dF Galaxy Redshift Survey
combine with the incredible recent progress in CMB
data to show spectacularly good agreement with a
`standard model' for structure formation. This consists
of a scalar-mode adiabatic CDM universe with scale-invariant
fluctuations. Measuring the exact parameters of this
model is rendered difficult by the intrinsic degeneracies
of the structure-formation process, but  progress is
being made. The most recent data yield
$\Omega_m = 0.25 \pm 15\%$ and $h=0.73 \pm 5\%$; these
figures accord well with independent constraints, and it
is very hard to believe that they are incorrect.

Allowing extra degrees of freedom, such as massive neutrinos,
vacuum equation of state $w\ne 1$, or extra relativistic content
worsens the agreement with independent constraints on
$h$ and $\Omega_m$. This both supports the simplest picture
and allows us to set interesting limits on these 
non-standard ingredients.

For the future, we can look with anticipation to 
meaningful tests of inflation: the current data are consistent
with $n=1$ to an error of $\pm 0.03$. Once this is halved,
plausible levels of tilt will come within our sensitivity.
The tensor fraction is a less clear target, but the motivation
to improve on the current weak upper limits will remain strong.

It should of course not be forgotten that the large-scale structure
we measure locally consists of galaxies. In this paper,
the physics of galaxy formation has been sadly ignored,
but this will be the increasing focus of LSS studies:
not just the global parameters of the universe, but the
detailed understanding of how the complex structures
around us formed.

\section*{Acknowledgements}

This paper has drawn on the body of results achieved by my colleagues
in the 2dF Galaxy Redshift Survey team: Matthew Colless (ANU), 
Ivan Baldry (JHU), Carlton  Baugh (Durham), 
Joss Bland-Hawthorn (AAO), Terry Bridges (AAO),
Russell Cannon (AAO), Shaun Cole (Durham), 
Chris Collins (LJMU), Warrick Couch (UNSW), 
Gavin Dalton (Oxford), Roberto De Propris (UNSW), 
Simon Driver (St Andrews), George Efstathiou (IoA), 
Richard  Ellis (Caltech), Carlos Frenk (Durham), 
Karl Glazebrook (JHU), Carole Jackson (ANU), Ofer Lahav (IoA), Ian Lewis (AAO), 
Stuart Lumsden (Leeds), Steve Maddox (Nottingham), Darren Madgwick (IoA), 
Peder Norberg (Durham), Will Percival (ROE), 
Bruce Peterson (ANU), Will Sutherland (ROE), Keith Taylor (Caltech).
The 2dF Galaxy Redshift Survey
was made possible by the dedicated efforts of the staff
of the Anglo-Australian Observatory, both in creating the 2dF
instrument, and in supporting it on the telescope.


\section*{References}

\japref Baugh C.M., Efstathiou G., 1993, MNRAS, 265, 145
\japref Baugh C.M., Efstathiou G., 1994, MNRAS, 267, 323
\japref Bond J.R., Szalay A., 1983, ApJ, 274, 443
\japref Bond J.R., Efstathiou G., Tegmark M., 1997, MNRAS, 291, L33
\japref Bucher M., Moodley K., Turok, N., 2002, Phys. Rev. D, 66, 023528
\japref Burles S., Nollett K.M., Turner M.S., 2001, ApJ, 552, L1
\japref Cole S., Hatton S., Weinberg D.H., Frenk C.S., 1998, MNRAS, 300, 945
\japref Colless M. et al., 2001, MNRAS, 328, 1039
\japref Efstathiou G., Bond J.R., 1999, MNRAS, 304, 75
\japref Efstathiou G. et al., 2002, MNRAS, 330, L29
\japref Elgaroy O. et al., 2002, Phys. Rev. Lett., 89, 061301
\japref Evrard A., 1997, MNRAS, 292, 289
\japref Feldman H.A., Kaiser N., Peacock J.A., 1994, ApJ, 426, 23
\japref Freedman W.L. et al., 2001, ApJ, 553, 47 
\japref Garnavich P.M. et al., 1998, ApJ, 509, 74
\japref Goldberg D.M., Strauss M., 1998, ApJ, 495, 29
\japref Hambly N.C., Irwin M.J., MacGillivray H.T., 2001, MNRAS, 326 1295
\japref Hu W., Sugiyama N., 1995, ApJ, 444, 489
\japref Hu W., Fukugita M., Zaldarriaga M., Tegmark M., 2001, ApJ, 549, 669
\japref Lahav O. et al., 2002, MNRAS, 333, 961
\japref Lewis I.J. et al., 2002, MNRAS, 333, 279
\japref Loeb A., Barkana R., 2001, ARAA, 39, 19
\japref Maddox S.J., Efstathiou G., Sutherland W.J., Loveday J., 1990a, MNRAS, 242, 43{\sc p}
\japref Maddox S.J., Sutherland W.J., Efstathiou G., Loveday J., 1990b, MNRAS, 243, 692
\japref Maddox S.J., Efstathiou G., Sutherland W.J., 1990c, MNRAS, 246, 433
\japref Mann R.G., Peacock J.A., Heavens A.F., 1998, MNRAS, 293, 209
\japref Meiksin A.A., White M., Peacock J.A., 1999, MNRAS, 304, 851
\japref Mould J.R. et al., 2000, ApJ, 529, 786
\japref Narayanan V.K., Spergel D.N., Dav\'e R., Ma C.-P., 2000, ApJ, 543, L103
\japref Norberg P. et al., 2001, MNRAS, 328, 64
\japref Percival W.J. et al., 2001, MNRAS, 327, 1297
\japref Percival W.J. et al., 2002, MNRAS, 337, 1068
\japref Schlegel D.J., Finkbeiner D.P., Davis M., 1998, ApJ, 500, 525
\japref Seljak U., Zaldarriaga M., 1996, ApJ, 469, 437
\japref Spergel D.N. et al., 2003, astro-ph/0302209
\japref Stoughton C.L. et al., 2002, AJ, 123, 485
\japref Verde L. et al., 2002, MNRAS, 335, 432
\japref Vilenkin A., 2001, hep-th/0106083
\japref Vittorio N., Silk J., 1985, ApJ, 297, L1
\japref Zaldarriaga M., Spergel, D., Seljak U., 1997, ApJ, 488, 1

\end{document}